\documentclass[12pt]{article}
\usepackage{amsmath}
\usepackage{amsfonts}
\usepackage{epsfig}
\newcommand{\beq}{\begin{equation}}
\newcommand{\eeq}{\end{equation}}

\newcommand{\pd}{\partial}
\newcommand{\edc}{\end{document}}

\newcommand{\by}{\bar {y}}
\newcommand{\brho}{\bar{\rho}}

\title{\bf Entropy in Toy Regge models}
\author{M.A. Braun\\
Dept. of High Energy physics,
Saint-Petersburg State University,\\
198504 S.Petersburg, Russia}

\begin{document}
\maketitle
\begin{abstract}
The probabilistic interpretation of  the standard Regge-Gribov model with triple pomeron interactions is discussed.
It is stated that
introduction of probabilities within this model is not unique
and depends on what is meant under the relevant substructures,
The traditional interpretation in terms of partons (quarks and gluons) is shown to be external to the model,
imported from the QCD, and actually referring to the single pomeron exchange without interactions. So this interpretation
actually forgets the model as such. Alternative probabilities based on the pomerons as basic quantities within the model are discussed.
Two different approaches are considered, based either on the pomerons in Fock's expansion of the wave function or on pomeron propagators
in Feynman diagrams. These pomeron probabilities and entropy turn out to be very different from the mentioned standard ones in the purely
probabilistic treatment. The entropy, in particular, either rises with the rapidity and saturates at a certain fixed value or
first rises, reaches some maximum and goes down to zero afterwards.
Possible observable manifestations of these probabilities and entropy are to be seen in the
distributions of the cross-section in powers $n$ assuming that their dependence of the coupling constants $g$ to the participants
is presented as a series in $g^n$.
\end{abstract}

\section{Introduction}

In the studies of the high-energy behavior of the amplitudes generated by strong interactions a notable place has been taken by
the Regge-Gribov models  based on the exchange of local  pomerons evolving in rapidity and
self-interacting with a non-Hermithean Hamiltonian. Being essentially a  semi-phenomenological approach, such models describe well
the bulk features of high-energy interactions with certain well-defined phenomenological parameters, such as the intercept and slope of the pomeron and
strength of its imaginary self-coupling. Defined initially in the two-dimensional transverse world they  unfortunately cannot be solved exactly
beyond  the quasi-classical approximation or in the renormalization group approximation in the vicinity of perturbatively found fixed points.
For this reason much attention was given in the past to the approximation of zero slope "Toy models", in which  the models reduce to fields depending only on rapidity
("zero-dimensional" in the transverse variables). In this approximation the models actually transform to one dimensional quantum mechanical systems
living in the imaginary time $t=-iy$, where $y$ is the rapidity, and a peculiar non-hermithean Hamiltonian symmetric in $p$ an $q$.

Some years ago it was proposed to look at toy Regge-Gribov models (TRGM)  in the framework of the so-called statistical reaction-diffusion
approach ~\cite{redif}. In this approach the system was described by a set of probabilities $P_n$ for the  creation and decay of a given number
$n$ of quasi-particles (e.g. pomerons).
In fact much earlier this probabilistic approach was developed in the context of dipole scattering in the QCD as a toy model for dipole-dipole scattering
in the approximation of the so called "big loops", which describes fan diagrams attached to the projectile and inverse fan diagrams attached to the target
joined at some middle rapidity $y_0$ between the rapidities $y$ and zero of the projectile and target ~\cite {musal}. it was formulated exclusively
in terms of probabilities $P_n(y)$ with  a particular evolution equation in  $y$. The total probability $P_n$ was presented as a convolution of
probabilities for the joining fan contributions, schematically
\beq
P=P(y_0)\otimes P(y-y_0).\label{bl}\eeq
On physical grounds (projectile-target symmetry) one would prefer $y_0=y/2$ and with this choice (\ref{bl}) was used in several applications
~\cite{musal,bravac}. However in fact (\ref{bl}) does depend on $y_0$ except in the limit $y\to\infty$when it corresponds to the unitarity in the $t$-channel
in the BFKL framework.
Since then much efforts have been directed to
achieve independence of this transitional rapidity $y_0$, which concentrated on the change  of the probabilities $P_n$ within this approach
~\cite{musal,1,2,3,4,5}.
Note that in the Regge-Gribov field theoretical approach this independence (or "$t$-channel unitarity") is automatic, since it simply corresponds to the
trivial relation
\[e^{-Hy}=e^{-Hy_0}e^{-H(y-y_0)}\]
where $H$ is the Hamiltonian.
So attempts to restore $y_0$ independence at finite $y$ in the probabilistic approach inevitably lead to field models with Hamiltonians
much more complicated than in TRGM, with inclusion of infinite number of
different multipomeron vertices. In this paper we restrict ourselves to the standard TRGM with only triple pomeron vertices and so leave more sophisticated models
outside our attention.

Returning to TRGM, in ~\cite{redif} it was noted that the standard one with only a triple pomeron interaction does not admit a simple probabilistic interpretation since
the probabilities $P_n$ introduced in the analogy with the statistical approach do not preserve positivity with the growth of rapidity.
However in ~\cite{bravac} it was noted that if only splitting (or merging) vertices are retained in the Hamiltonian, when the amplitude reduces to
 the fan diagram
approximation, after some manipulations one  returns to the above discussed probabilistic models with well-defined probabilities $P_n$.

In this note we try to reconsider the probabilistic interpretation of the standard TRGM. We note that the definition of
probabilities $P_n$ in the model is far from being unique and depends on what is taken to be the basic substructures.
In the standard probabilistic approach they are partons borrowed from the QCD (quarks and gluons).
So the corresponding probabilities and entropy can be called partonic.
However staying strictly within TRGM we do not see any partons. Rather the relevant basic quantities are pomrerons characterized by
Fock's representation of the wave function $\Psi(y)$
\beq
\Psi(y)=\sum_nc_n(y)\Psi_n.\label{fok}\eeq
where $\Psi_n$ are states with $n$ pomerons. Unlike the standard Quantum Mechanics
$\Psi(y)$ has a norm which is not definite positive.
Obviously coefficients $c_n$ in some way characterize the relative contribution of $n$ pomerons at rapidity $y$.
Based on these considerations one can try to introduce probabilities $P_n$ to find $n$ pomerons in the system.
These pomeronic probabilities will be different from the partonic ones introduced in the probabilistic approach.

In TRGM one can pass to real wave functions and Hamiltonian.
In the simplified quasi-classical approximation,  when the amplitude can be found analytically,
one finds that $c_n(y)$ can be both positive and negative and $|c_n(y)|$
can be greater than unity (and in fact grow indefinitely as $y\to\infty$). So
 it is impossible to directly relate $c_n$  to some probabilities.
 However there is a simple (but certainly not unique) way out. Having in mind that $|c_n(y)|$ grow with $n$ and $y$
 one can define the probabilities as
 \beq
 P_n(y)=\frac{|c_n(y)|}{R^n(y)},\ \ R(y)>0
\label{pn}
\eeq
and search for $R(y)$ from the requirement
\beq
Z(y)=\sum_{n=1}P_n(y)=1
\label{zy}
\eeq
With so  found $R(y)$ probabilities  $P_n$ satisfy all requirements for  probabilities
\[P_n(y)\geq 0,\ \ \sum_nP_n(y)=1\]
and correctly describe the relative contribution of a given number $n$ of pomerons.
So they can be safely taken as the desired  pomeronic probabilities.
The weak point of this definition is the possibility to solve (\ref{zy}) for a positive $R(y)$.
However as we shall see this obstacle does not arise.

We shall see  in Section 3.2 that for the case when there occurs only splitting (or merging) of pomerons these probabilities are
quite different from those which figure in the common probabilistic models. They  have a completely different
behavior at large $y$ and a different entropy.

The described method is not the only one which can be used to introduce the probability to nave $n$ pomerons.
In fact this depends on what one defines as an $n$-pomeron state. Some alternative (although with a limited application)
will be discussed in Section 3.3. It will lead to still more different behavior of the probabilities and entropy.

\section{Standard TRGM}
In this section we recall the well-known facts about the zero-dimensional ("Toy") standard Regge-Gribov model.
The "state' in the model is characterized by the wave function $\Psi(y)$ depending on rapidity $y$ and evolving in $y$
according to the quasi-Schroedinger equation
\beq
\frac{d\Psi(y)}{dy}=-H\Psi(y)
\label{scheq}
\eeq
where $H$ is the non-Hermithean Hamiltonian depending on the operators $\psi(y)$ and $\psi^\dagger(y)$ for annihilation and creation of pomerons.
After transformation to new operators
\[\phi^{\dagger}=-iu,\ \ \phi=-iv\]
the Hamiltonian becomes real
\beq
H=-\mu uv+\lambda uv(u+v).\label{ham}
\eeq
The operators $u$ and $v$ are subject to the abnormal commutation relation
\beq
[v,u]=-1.
\label {com rel}
\eeq
The vacuum state satisfies
$v\Psi_0=0$.
All other states are obtained by application of $u$ on the vacuum. In particular
\[\Psi_n=u^n\Psi_0\]
is a state with $n$ pomerons.
In the $u$-representation, when $u$ is the operator of multiplication,
\[v=-\frac{\pd}{\pd u}\]
so that the Schroedinger equation (\ref{scheq}) becomes an equation in partial derivatives in $y$ and $u$.

This equation can be technically  solved either by expanding in the eigenstates $\Psi^{(n)}$ of the Hamiltonian
or by direct evolution by, say, the Runge-Kutta method, as in ~\cite{bravac}. In the free theory ($\lambda=0$) the energy spectrum
is evidently $E_n=-\mu_n$, $n=0,1,2,...$, so that with $\mu>0$ (supercritical pomeron)  all non-trivial solutions
grow indefinitely as $y\to\infty$. However, as was shown long ago, the interaction radically changes this behavior
showing the decisive influence of quantum corrections (loops). As was established in ~\cite{jengo} at small $\lambda/\mu$
the ground state takes a small positive value
\beq
E_0=\frac{\mu \rho}{\sqrt{2\pi}}e^{-\rho^2 /2},\ \ \rho=\frac{\mu}{\lambda}
\label{e0}
\eeq
and the wave function slowly goes to zero at $y\to \infty$.

In the direct evolution one uses the differential equation (in the $u$-representation)
\beq
\frac {\pd \Psi(y,u)}{\pd y}=-\Big(\mu u\frac{\pd}{\pd u}-\lambda u^2\frac{\pd}{\pd u}
+\lambda u\frac{\pd^2}{(\pd u)^2}\Big)\Psi(y,u)
\label{difeq}
\eeq
with the initial condition, taken typically in the eikonal approximation
\beq
\Psi(0,u)=1-e^{-g_1u}
\label {psi0}
\eeq
where $g_1$ is the pomeron-nucleon coupling  in the projectile
The scattering AA amplitude is then given as ~\cite{bravac}
\beq
{\rm Im}\,{\cal A}(y)=\Psi(y, g_2)
\label{amp}
\eeq
where $g_2$ is the pomeron-nucleon coupling in the target at rapidity $y$..

This well-known information by itself however tells us nothing about the statistical characteristics of the model.

\section{The quasi-classical approximation. Fan diagrams}
In the quasi-classical approximation the scattering amplitude reduces to a set of tree diagrams
shown in Fig. 1 $a$ for hA scattering and Fig. 1 $b$ for AA scattering.
\begin{figure}
\begin{center}
\epsfig{file=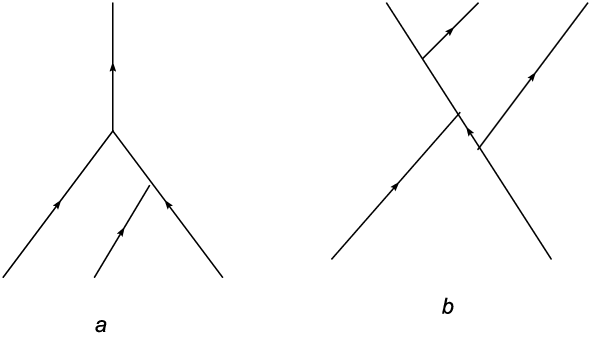, width=8 cm}
\caption{Tree diagrams for proton-nucleus (a) and nucleus-nucleus (b) scattering}
\end{center}
\label{fig1}
\end{figure}
In this approximation
the standard TRGM was studied by D.Amati {\it et al.} long ago ~\cite{amati}.
They searched for the solution of the
 equations of motion
\[ \dot{u}=\frac{\partial u}{\partial y}=\frac{\pd H}{\partial v}=-\mu u+\lambda(2uv+u^2),\]
\beq\dot{v}=\frac{\partial v}{\partial y}=-\frac{\pd H}{\partial u}=\mu v-\lambda(2uv+v^2).\label{eqm}\eeq
under boundary conditions
 \beq
 u(Y)=g_2,\ \ v(0)=g_1\label{bc}\eeq
 where $Y$ is the rapidity of the target with the projectile at rest
 and $g_2$ and $g_1$ are the couplings to external sources.
 The physical quantity of interest was action $\cal A$
 \beq
 {\cal A}
 =\lambda\int_0^Ydy v^2u+g_1u(0),
 \label{action}\eeq
 which determined the scattering matrix.
 However they were not able to find the solution explicitly at all rapidities but only
 in the limit $Y>>1$.

\subsection{Reaction-diffusion probabilities}

An especially simple case  occurs when  field $u$ couples only once to the projectile (nucleon). Then one can drop
the term with the product $uv$ in the first equation (\ref{eqm}) and obtain the evolution equation for $u$ only.
The corresponding amplitude  is then given by a set of fan diagrams propagating from the projectile (nucleon) to the target (nucleus).
Fig. 1 $a$. and can be found explicitly ~\cite{schwimmer}.
The wave function $\Psi(y,u)$ in the $u$ representation is then \cite{bravac}
\beq
\Psi(y,u)=\frac{g_1u e^{\mu y}}{1+\frac{\lambda u}{\mu}(e^{\mu y}-1)}.
\label{f}
\eeq

Having this simple and explicitly known solution one may try
to interpret the amplitude in terms of probabilities.
In  ~\cite{redif} the fan case was interpreted as a particular case of
a  reaction-diffusion process. Following this approach the probabilities to have $n$ partons
were introduced in ~\cite{bravac} as follows.
Following (\ref{fok}) present
$\Psi(y,u)$ as a power series in $u$
\beq
\Psi(y,u)=\sum_{n=1}c_n(y)u^n.
\label{powerexp}
\eeq
The evolution equation for $\Psi$ generates a system of equations for coefficients
$c_n(y)$:
\beq
\frac {dc_n(y)}{dy}=\mu n c_n(y)-\lambda (n-1)c_{n-1}(y),\ \ c_{-1}(y)=0.
\label{cny}
\eeq
Rescale $y=\bar{y}/\mu$ and present
\beq
c_n(\bar{y})=\frac{1}{n!}\Big(-\frac{\mu}{\lambda}\Big)^{1-n}\nu_n(\bar{y})
\label{cn}
\eeq
to obtain an equation for $\nu_n$
\beq
\frac{d\nu_n(\bar{y})}{d\bar{y}}= n
\nu_n(\bar{y})+n(n-1)\nu_{n-1}(\bar{y}).
\label{nuevol}
\eeq
Note that this equation does not depend on $\mu$ nor on $\lambda$.
Now introduce probability $P_n$ by the relation
\beq
\nu_n(\bar{y})=\sum_{k=n}^\infty P_k(\bar{y})\frac{k!}{(k-n)!}
\label{nun}
\eeq
with the inverse relation
\beq
P_n(\bar{y})=\sum_{k=n}^\infty(-1)^{k-n}\frac{\nu_k(\bar{y})}{n!(k-n)!}.
\label{rhonu}
\eeq
One finds that $P_n$ satisfy the evolution equation in $\bar{y}$
\beq
\dot{P}_n(\bar{y})=-nP_n(\bar{y})+(n-1)P_{n-1}(\bar{y}).
\label{prob}
\eeq
With $P_n(0)=\delta_{n1}$  the solution is
\beq
P_n(y)=e^{-\mu y}a^{n-1},\ \ a=1-e^{-\mu y}
\label{rhon}
\eeq
and the property
\beq
\sum_{n=1}P_n(y)=1\ \ {\rm at\ all\ rapidities}.
\eeq
So  taking into account that $P_n$ are positive  one can assume that $P_n$ give some  probabilities.
Precisely the same probabilities were introduced in previous papers in the probabilistic approach (see e.g.  ~\cite{kharzeev}).
Considering the  model as a simplified QCD the probabilities have been interpreted as the ones to have
$n$ partons (quarks and gluons) in the system.
At $y>>1$ one finds $a\simeq 1$ and the probabilities do not depend on $n$.
This fact was interpreted  in~~\cite{kharzeev}  as a manifestation of achieving the maximally  entangled state and "information scrambling"
in terms of the information theory.

The entropy following from the probabilities (\ref{rhon}) is
\beq
S(y)=\mu y-e^{\mu y} a\ln a.
\label{sk}
\eeq
At large  $y$ this entropy grows linearly $S(y)\sim \mu y$. It is shown in Fig. 2 by the upper curve.
From (\ref{prob}) it follows that the average $\bar{n}$ satisfies a simple equation
\beq
\frac{d\bar{n}(y)}{dy}=\mu \bar{n},\ \ \bar{n}(y)=\sum_{n=1}nP_n(y)
\label{avny}
\eeq
with the general solution
\beq
\bar{n}(y)=Ce^{\mu y}.\label{avn}\eeq
Note that neither $P_n$ nor $\bar{n}$ depend on the pomeron coupling $\lambda$ in TRGM.
With $P_n(0)=\delta_{n1}$ one has
$
\bar{n}(y)=e^{-\mu y}.
$
In the spirit of the parton-hadron duality the emitted partons have been related to emitted hadrons,
so that the probabilities and entropy have been related to observed hadron multiplicities. In this way
the entropy ({\ref{sk}) has been treated as an observable quantity to be found from the experimental data
\cite{hentch1,hentch2}.

Staying within  TRGM these results are easily interpreted. Due to the AGK cancelations
~\cite{agk} in the quasi-classical approximation inclusive cross-section coming from all nontrivial
diagrams, like shown in Fig 1, cancel and all contribution comes exclusively from the zero-order diagram, that is, the single pomeron exchange,
as $Ce^{-\mu y}$ where $C$ depends on the sort of the emitted particle. Of course it does not depend on $\lambda$ being the zero order term in the
perturbative expansion. This agrees with the average $\bar{n}$ interpreted as
the average number of emitted hadrons in the probabilistic approach. TRGM by itself does not know about the inner structure of the pomeron exchange,
so the probabilities $P_n$ are  external to the model supplemented to it by the relation to the QCD.
The transformations (\ref{cn}) to (\ref{prob}) proposed in ~\cite{redif,bravac} serve to artificially relate $c_n$ and $P_n$ and in this way
they  liquidate TGRM almost completely leaving only the zero order term.the single pomeron exchange.
The net effect of these transformations seems to impose the AGK cancelations on the model.

It is important to note that these partonic $P_n$  do not refer directly to the number of pomerons in the Fock expansion of the wave function
(\ref{powerexp}). Therefore one can alternatively study probabilities of a different sort, describing the structure of the system not in terms
of partons absent in the system
and artificially brought by analogy with the QCD but  rather in terms of pomerons, which are in fact
actual components of the system.
Possibilities to introduce such pomeronic probabilities and entropy will be considered in the next chapters.

\begin{figure}
\begin{center}
\epsfig{file=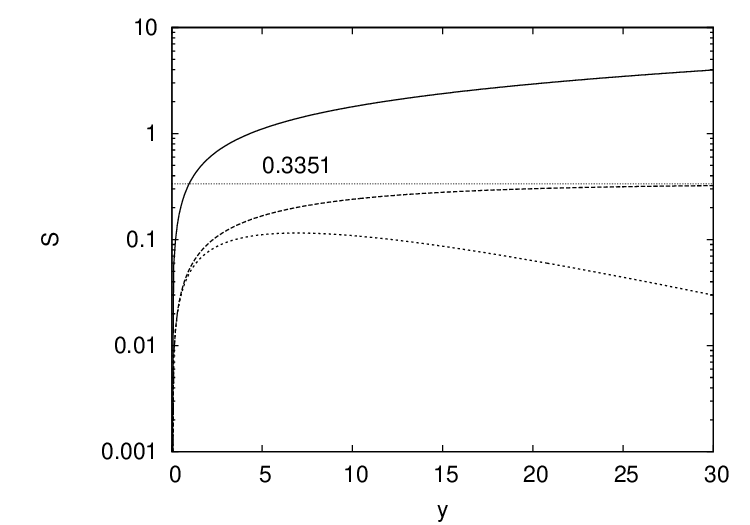, width=8 cm}
\caption{Entropy for the fan amplitude with $\mu=0.1$ and  $\lambda =0.01$ as  a function
of rapidity. The upper curve shows entropy (\ref{sk}) from the reaction-diffusion approach. The middle curve
shows entropy (\ref{sresc}) from the rescaling approach. The lower curve shows entropy (\ref{ss1}) from the diagrammatic
approach}
\end{center}
\label{fig2}
\end{figure}

\subsection{Number of pomerons from $c_n(y)$. Rescaling method.}.
As stated in the Introduction ne can define the number of pomerons $n$ directly from the expansion (\ref{powerexp}) as the power of $u$.
adequately rescaling  coefficients $c_n(y)$ and defining the probabilities as
\beq
P_n(y)=|c_n(y)|R^{-n}(y),
\label{rhoresc}
\eeq
where $R(y$ is found as a positive  solution of the equation
\beq
Z(R)=\sum_{n=1}\frac{|c_n|}{R^n}=1.
\label{norm}
\eeq
Actually this procedure is to rescale positive $|c_n(y)|$ by factor $P^n$ to obtain probabilities with the necessary properties

Let us see what this rescaling method gives for  fan diagrams. From (\ref{f}) putting $g_1=1$ we find
\beq
|c_n|=\frac{\rho}{a}\Big(\frac{a}{\rho (1-a)}\Big)^n,\ \ \rho=\frac{\mu}{\lambda}.
\label{c1}
\eeq
Correspondingly the probabilities are
\beq
P_n=\frac{\rho}{a}\Big(\frac{a}{\rho R(1-a)}\Big)^n
\eeq
with the norm
\beq
Z(R)=\frac{\rho}{a}\sum_{n=1}\Big(\frac{a}{\rho R(1-a)}\Big)^n=\frac{1}{R(1-a)-a/\rho}.
\eeq
Condition $Z(R)=1$ determines the contribution from a single pomeron exchange
\beq
R=\frac{a+\rho}{\rho (1-a)}.
\eeq
With this $R$ the probabilities determined from (\ref{rhoresc}) are found to be
\beq
P_n=\frac{\rho}{a}\Big(\frac{a}{a+\rho}\Big)^n.
\label{rofan1}
\eeq
At $y=0$ we have $a\to 0$ and $P_n(y=0)=\delta_{n1}$ as expected.  At $y>0$ the probabilities are falling win $n$
as powers
\[ P_{n+1}=P_n\frac{a}{a+\rho}.\]
In the high-energy limit ($a\to 1$) we obtain $P_n\sim (1+\rho)^{-n}$.
The average number of pomerons is
\beq
\bar{n}=\sum_{n=1}nP_n=\frac{\rho+a}{\rho}.
\eeq
and
the corresponding entropy is given by
\[
S=-\frac{\rho}{a}\sum_{n=1}
\Big(z^n\ln\frac{\rho}{a}+nz^n\ln z\Big),\ \ z=\frac{a}{a+\rho},
\]
and after some calculations
\beq
S=\frac{a+\rho}{\rho}\ln (a+\rho)-\frac{a}{\rho}\ln a -\ln \rho
=\bar{n}\ln\bar{n}-(\bar{n}-1)\ln(\bar{n}-1).
\label{sresc}
\eeq
At $y=0$ we get $\bar{n}=1$ and $S=0$. With the growing $y$ both the average number of pomerons and entropy $S(y)$ grow monotonously
until finite limits:
\beq
\bar{n}_\infty=1+\frac{1}{\rho},\ \ S_{\infty}=\frac{1+\rho}{\rho}\ln(1+\rho)-\ln \rho.
\label {sresc1}
\eeq
For the free theory $\rho\to\infty$ we correctly get $P_n=\delta_{n1}$ and $S=0$.
This entropy is shown in Fig. 2 by the middle curve. The horizontal line marks its limit $S_\infty=0.3551$.

One may ask if these pomeronic probabilities and entropy are observable. The answer is  yes. According to (\ref{amp})
the scattering amplitude is obtained after substitution $u\to g_2$ in $\Psi(y,u)$ where $g_2$ is the coupling to the
nucleon in the target. So distribution in $n$ means the distribution in powers $n$ of the dependence of the amplitude
on $g_2$. So one can observe $n$ by studying the scattering on the target with different values of $g_2$, which
corresponds to different spatial interaction points governed by the transverse distribution $T_A(b)$ in the nucleus target.

This definition of probabilities has the advantage of  being directly based on the number $n$ of pomerons in the development
(\ref{powerexp}) of the wave function in the fan diagram approximation. However
in the quasi-classical approximation it is technically  very difficult to introduce it for more complicated diagrams like Fig. 1 $b$, when the
solution cannot be obtained in the form (\ref{powerexp}) but only from the quasi-classical equations of motion.
A new definition of pomeronic probabilities valid for any diagrams in the quasi-classical approximation will be presented in the
next chapter.

\subsection{Number of pomerons from the diagrams}
To define the number of pomerons one may forget about the expansion (\ref{powerexp}) and consider instead   the amplitude
as a set of Feynman diagrams.
One may define the number of pomerons $n$ as the number of pomeron propagators in the diagram.
It may correspond to the bare pomeron, that is simply the line, or the 'physical' or 'full' pomeron, that is the full
two-reggeon Green function consisting of the bare line with all loop insertions.
The problem with the second choice is that it is very difficult to locate physical pomerons in the general
way. To do that one has to consider skeleton diagrams made of full pomerons, each of them containing an infinite number
of bare pomerons. We do not know any technique which allows to find the number of full pomerons in a diagram in any general way.
But this problem disappears once we restrict ourselves
 to the quasi-classical approximation neglecting the loops. Then the physical pomeron
becomes identical with the bare one and the number of pomerons is just the total number of lines.
Of course this definition of the number of pomerons is quite different from their definitions $n$ in the expansion of the wave function
(\ref{powerexp}). But it has a clear physical meaning and, which is important in the quasi-classical picture, can be applied to the tree-diagrams
of any structure, including both Fig. 1 $a$ and $b$.

Let a  Feynman diagram  contain
$E$ external lines $I$ internal lines and $V$ vertices.
With the triple interaction relevant for  Regge models one has the relation
$
2I+E=3V$. The number of loops $L$ in a connected
diagram is
$
L=I-V+1.$
In the quasi-classical approximation we take the number $n$ of pomerons as
\beq
n=E+I
\label{ndiag}
\eeq
with  $L=0$. Then
 the number of pomerons can be uniquely expressed via $E$ or via $V$
\beq
n=E+V-1=2E-3=2V+1.\label{nop}
\eeq
Note that $n$ is odd, so that with this choice probabilities to find an even number of pomerons are zero $P_{2N}=0$, $N=0,1,2...$

In the quasi-classical approximation only tree diagrams survive shown in Fig. 1 $a$ for hA scattering and Fig. 1 $b$ for AA scattering.
If the couplings to the projectile and target are $g_1$ and $g_2$ respectively, then the contribution with $E=N_1+N_2$ external lines
will be proportional to $g_1^{N_1}g_q^{N_2}$. To simplify we consider a simple case $g_1=g_2=g$ (identical nuclei). Then the contribution
from $n$ pomerons will be proportional to $g^N$ where $n=2N-3$.
The pomerons with this new definition are different from those in Fock's expansion (\ref{powerexp}) used in the previous approach.
They are counted in a different manner. Say for the diagram in Fig 1 $a$ the number of external lines is N= 4 and so the number of pomerons
determined from the diagram is $n= 5$, although the latter comes from the term $u^3$ in (\ref{powerexp}) and the number of pomerons in the
previous definition is only 3.

To pass to probabilities we present ${\cal A}$
 as a series in powers of $g$
\beq
{\cal A}=\sum_{N=2}(-1)^N g^NA_N,\ \ N=1,2,...,\ \ n=2N-3.
\label{expa}\eeq
So the number of pomerons is always odd and we take into account that the sign of contributions is
alternating with $N$.
We are going to present $A_N$ as a product of the probability $P_n$ to have $n$ pomerons multiplied by the contribution to
$A$ from these $n$ pomerons $R^n$, that is, as
\beq
A_N=R^nP_n,\ \ n=2N-3
\label{ann}\eeq
where $P_n$ are positive and properly normalized
\beq
\sum_{N=2}P_n=1,\ \ n=2N-3.
\label{norm}\eeq

Using the scaling procedure from the previous section we
construct
\[X_N(R)=\frac{A_N}{R^n},\ \ n=2N-3\]
and the norm $Z(R)$ depending on $R$
\beq
Z(R)=\sum_{N=2}X_N(R),\ \ n=2N-3.\label{np}\eeq
obeying
$Z(R)$.=
Then we take
\beq
P_n=X_N(R)=\frac{A_N}{R^n},\ \ n=2N-3
\label{rho}\eeq
and have
\[A_N=R^n\frac{A_N}{R^n}=P_n R^n,\ \ n=2N-3\]
as desired with the properly normalized $P_n$.

Let us see how this technique works for fan diagrams.
In this case taking the coupling to both proton an nucleus equal to $g$ we have the
amplitude
\beq
{\cal A}= \frac{g^2e^{\mu y}}{1+\frac{g}{\rho}(e^{\mu y}-1)}=\frac{g^2}{1-a-ga/\rho}
\label{afan}
\eeq
where $\rho=\mu/\lambda$ and $a$ was defined in (\ref{rhon}).
Developing in powers of $g$ we get in (\ref{expa})
\beq
A_N=(1-a)^{-N+1}\Big(\frac{a}{\rho}\Big)^{N-2}.
\eeq
So in our technique
\beq
X_N=(1-a)^{-N+1}\Big(\frac{a}{\rho}\Big)^{N-2}R^{-2N+3}\label{xn}\eeq
and
\[Z(R)=\frac{1}{R(1-a)-a/(\rho R)}\]
where we use
\beq\sum_{N=2}x^N=\frac{x^2}{1-x},\ \ \sum_{N=2}Nx^N=(2-x)\frac{x^2}{(1-x)^2}.
\label{sums}\eeq
The equation for $R$ is obtained as
\[R^2(1-a)-R-\frac{a}{\rho}=0\]
with a positive solution
\beq
R=\frac{1+\sqrt{1+4a(1-a)/\rho}}{2(1-a)}.
\label{pp}
\eeq

The probabilities turn out to be
\beq
P_n=\frac{\rho R(\rho R+a)}{a^2}\Big(\frac{a}{\rho R+a}\Big)^N,\ \ n=2N-3
\label{rgon1}
\eeq
The average number of pomerons $\bar{n}$ is given by
\beq
\bar{n}=2\bar{N}-3,\ \
\bar{N}=\frac{2\rho R+a}{\rho R}
\label{avbn}
\eeq
and the entropy is
\beq
S=\ln\frac{\rho R+a}{\rho R}+\frac{a}{\rho R}\ln \frac{\rho R+a}{a},
\label{ss1}
\eeq
where we once more used (\ref{sums}).

The found probabilities and entropy are radically different from what was obtained previously.
They depend on $\lambda$ via $\rho$ and as $\lambda\to 0$ they correctly go to $\delta_{n1}$.
They strongly diminish with the number of pomerons. Indeed we have
\beq
P_{n+2}=P_n\frac{a}{\rho R+a},\ \ n=1,3,5...\ \ {\rm with}\ \ P_1=\frac{\rho R}{\rho R+a}
\label{rhond}
\eeq
where the pomeron contribution is
\beq
R=ce^{\mu y},\ \ c=\frac{1}{2}\Big(1+\sqrt{1+4a(1-a)/\rho}\Big),\ \ 1<c<\frac{1}{2}(1+\sqrt{1+1/\rho}).
\label{pom}
\eeq
When rapidity grows also $R$ grows and the fall of $P_n$ becomes more and more
pronounced. At very high $y$ the probabilities return to their values at  $y=0$.
Correspondingly the entropy $S$ starts at $S=0$ then grows achieving its maximum at
approximately $\mu y=0.7$ for $\rho=1$ and slowly falls to zero afterwards, as illustrated in
Fig. 2 for $\mu=0.1$ and $\lambda=0.01$ by the lowest curve.

\section{Quasi-classical approximation. Multiple fans}
Staying within the approximation with only  splitting of pomerons one can find explicit amplitudes
for the case when the projectile interacts more than one time with the pomeron.
In the eikonal approximation this corresponds to taking expression (\ref{psi0}) for the initial amplitude.
Simple derivation then gives the wave function at any rapidity ~\cite{bravac}
\beq
\Psi(y,u)
=1-\exp \Big[-\frac{g_1ue^{\mu y}}{1+\frac{u}{\rho}\Big(e^{\mu y}-1\Big)}\Big].
\label{psi01}
\eeq
We recall that $\rho=\mu/\lambda$ and $g_1$ is the coupling of the pomeron to the projectile (nucleon).
Using this expression one can find coefficients $c_n(y)$ in the expansion (\ref{powerexp}) and from them
introduce the probabilities, either following the  reaction-diffusion approach passing as before
\[c_n(y)\to \nu_n(y)\to P_n(y)\]
or by our rescaling method via
\[P_n(y)=|c_n(y)|/R(y),\ \ \sum_{n=1}P_n(y)=1.\]
However explicit construction of $c_n(y)$ from (\ref{psi01}) is quite cumbersome and we prefer to act in a different way.
We shall calculate either $c_n$ or probabilities themselves at $y=0$ and then  evolve them to higher rapidities using the
evolution   equations.

\subsection{Reaction-diffusion probabilities}
 To find the probabilities one commonly introduces
 a generating function
\beq
\Phi(y,u)=\sum_{n=1}P_n(y) u^n
\label{phi}
\eeq
subject to the obvious conditions
\beq
\Phi(y,0)=0,\ \ \Phi(y,1)=1
\label{bound}
\eeq
following from its structure and normalization of $P_n$.
It has long been known that due to evolution equations (\ref{prob}) function $\Phi(y,u)$ obeys the equation
\beq
\frac{\pd \Phi(\by,u)}{\pd \by}=(u^2-u)\frac{\pd \Phi(\by,u)}{\pd u},\ \  \by=\mu y.
\label{eqphi}
\eeq
Its solution satisfying the boundary conditions (\ref{bound}) is ~\cite{bravac}
\beq
\Phi(\by,u)=\frac{1}{b}\Big(e^{-\brho(1-u)/(1-au)}-e^{-\brho}\Big)
\label{phiyu}
\eeq
where $\by=\mu y$, $a=1-\exp(-\by)$,  $\brho=g_1\rho$ and $b=1-\exp(-\brho)$.

For the eikonal wave function no term constant in $u$ is present, so that in (\ref{powerexp}) and (\ref{cn})
$c_0(y)=\nu_0(y)=0$  due to (\ref{cny}).
At $u=0$ we get from (\ref{nun})
\[P_0(y)+\Phi(y,1)=P_0(y)+1=0,
\]
which determines the nonphysical provability $P_0(y)$ to be equal to minus unity..

At $y=0$ we have $a=0$ and
\beq
\Phi(0,y)=\frac{1}{b}\Big(e^{\brho(1-u)}-e^{-\brho}\Big)
\label{phi0}
\eeq
From this expression  by multiple differentiation  we find the probabilities at $y=0$
\beq
P_n(0)=\frac{e^{-\brho}}{b}\,\frac{\brho^n}{n!},\ \ n\geq 1.
\label{probeik}
\eeq
The average $n$ at $y=0$ is found to be
$
\bar{n}(0)=\brho/b$..
From (\ref{avn}) it follows then that at all rapidities
\beq
\bar{n}(y)=\frac{\brho}{b}e^{\mu y}.
\label{avn51a}
\eeq
This corresponds to multiple  pomeron exchanges  in  eikonal diagrams.

Probabilities (\ref{probeik}) in turn determine coefficients $\nu_n(0)$ by (\ref{nun}). For $n\geq 1$
\[\nu_n(0)=\sum_{k=n}P_k(0)\frac{k!}{(k-n)!}
=\frac{e^{-\brho}}{b}\sum_{k=n}\frac{\brho^k}{(k-n)!}=\frac{1}{b}\brho^n\]
and coefficients $c_n(0)$ via $\nu_n(0)$ are found to be
\[c_n(0)=-(-1)^n\frac{1}{n!}\brho^{1-n}\frac{1}{b}\brho^n=-(-1)^n\frac{1}{n!}\frac{\brho}{b}.\]
The obtained $c_n(0)$ correspond to the eikonal wave function with a specific normalization
\beq
\Psi(0,u)=\frac{\brho}{b}\Big(1-e^{-g_1u}\Big).\ \ \brho=g_1\rho.
\label{psieik}
\eeq
As $g_1\to 0$ the normalization coefficient tends to unity and we get
\beq
\Psi(0,u)_{g_1\to 0}=g_1u,
\eeq
that is the initial state for a single fan. In this limit we find from (\ref{probeik}) $P_1=1$ and all $P_n$
for $n>1$ equal to zero as expected.

With many fans $\brho$ is different from zero and from (\ref {probeik}) one sees that the probabilities are distributed among many partons.
As  a result the entropy becomes greater than zero already at $y=0$ although the initial state corresponds to a given wave function.
This occurs because we fix the microstates as corresponding to a given number of partons, so that the initial state is rather an ensemble
of states with a different number of partons.

The probabilities at $y>0$ can be obtained from (\ref{probeik}) using evolution equations (\ref{prob}). We  cutoff the number of partons $n$ at
$n=N=300$. Below we present our numerical results for
our typical case $\mu=0.1$ and $\lambda=0.01$. We take $g_1=1$ so that $\brho=\rho=10$. In Fig. 3 we show the probabilities $P_n(y)$ for
$y=0,10,20,30$ ($\by=0,1,2,3$). For illustrative purposes we rescale the probabilities to have  equal maxima (around unity). One sees that
the spread in the number $n$ of pomerons grows with rapidity covering nearly all values $n<300$ at $y=30$
In Fig. 4 we show the entropy for $0<y<30$. As stressed  at $y=0$ the entropy is greater than zero
\[S(0)=2.561.\]
With the growth of $y$ the entropy grows practically linearly similar to the single fan case.
The abrupt  bending of the curve at $y>27$  does not reflect the actual behavior but only the result of a breakdown of
evolution due to accumulation of errors within the adopted precision.
\begin{figure}
\begin{center}
\epsfig{file=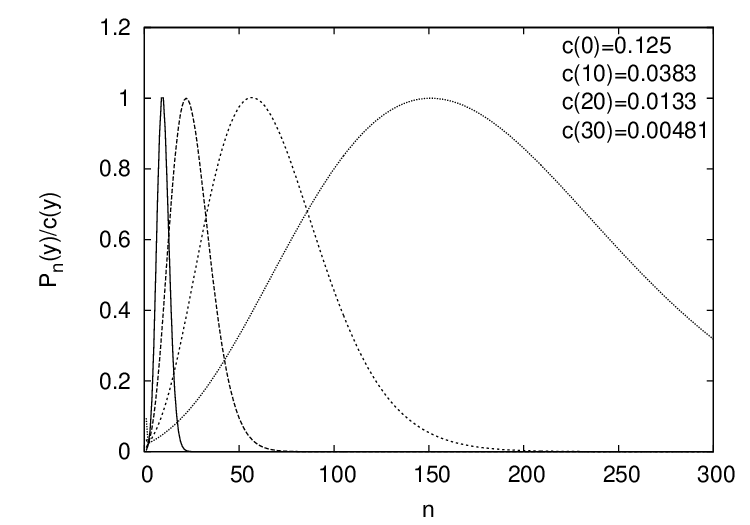, width=10 cm}
\caption{Probabilities  to have $n$ pomerons for eikonal  fans  in the reaction-diffusion approach for $0\leq n\leq 300$.
Shown are rescaled probabilities $P_n(y)/c(y)$ to achieve equal maxima of the curves, which correspond to $y=0,10,20,30$
from left to right. Values of $c(y)$ are correspondingly 0.125, 0.0383, 0.0133 and 0.00491.}
\end{center}
\label{fig3}
\end{figure}

\begin{figure}
\begin{center}
\epsfig{file=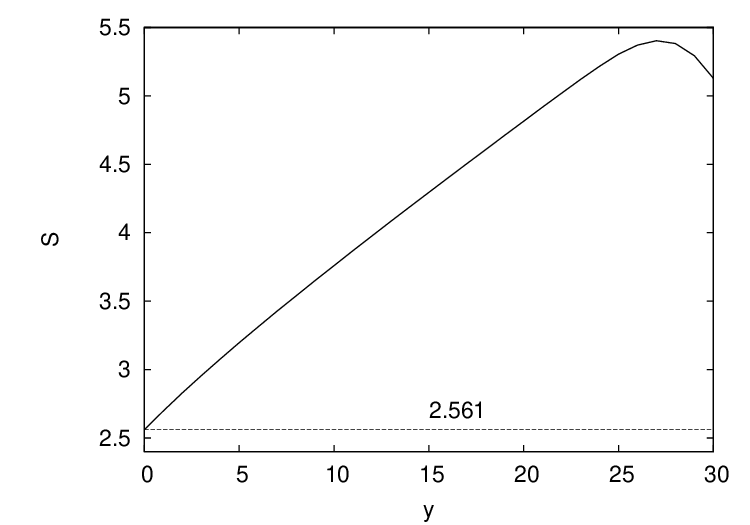, width=10 cm}
\caption{Entropy $S(y)$ for eikonal  fans  in the reaction-diffusion approach.
Bending of the curve at $y>27$ is only due to the evolution breakdown.}
\end{center}
\label{fig4}
\end{figure}

\subsection{The rescaling approach}
Contrary to the reaction-diffusion approach, which needs a particular normalization of the initial wave function,
the rescaling method can work with the original eikonal wave function for which at $y=0$ the coefficients in the expansion in powers $u^n$ are
$
c_n(0)=-(-1)^ng_1^n/n!$.
As before we put $g_1=1$. Starting from $c_n(0)$ we use Eqs. (\ref{cny}) and evolve $c_n(y)$ to higher rapidities.
Then, as described, we define $P_n$ according to Eqs. (\ref{rhoresc}) and (\ref{norm}).

In our numerical calculations we took $\mu=0.1$ and $\lambda=0.01$ as before. Eqs (\ref{cny}) were
cutoff at $n=N=80$. Our results for the entropy are shown in Fig. 5. in the left panel (the upper curve) and in the right panel
(practically a constant).
As before the entropy is greater than zero already at $y=0$
\[S(0)=0.8121.\]

\begin{figure}
\begin{center}
\epsfig{file=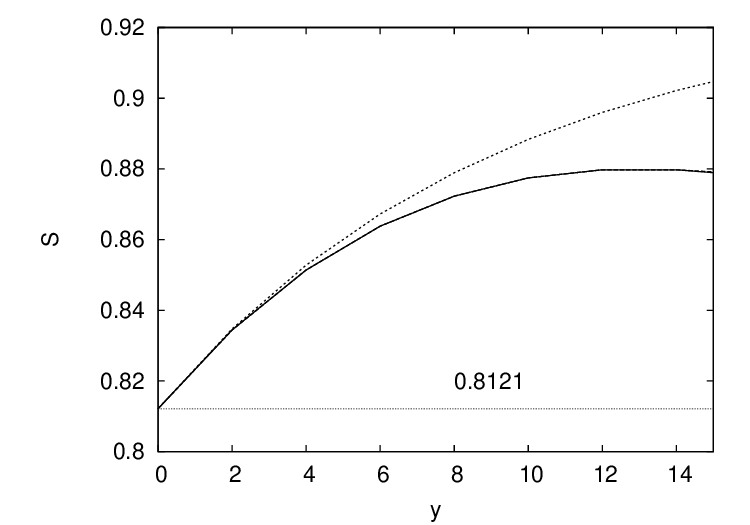, width=8 cm}
\epsfig{file=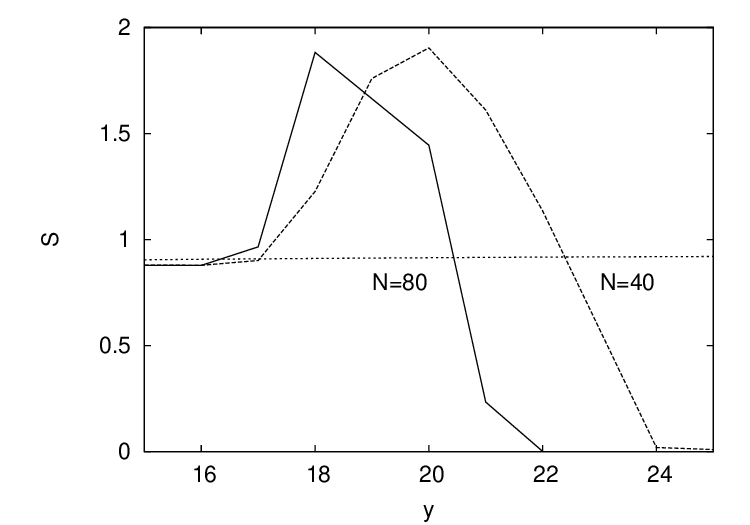, width=8 cm}
\caption{Entropy for AA scattering from the rescaling method with cutoff $n<N$ in the evolution equation for $P_n$
for the multiple fans and full quantum model. The results for the multiple fans are shown  in the left panel
by the upper curve and in the right panel by practically the constant curve, The results for the full quantum  case
for $y=20$ are practically independent of $N$ are shown in the left panel by the lower curve. The results  for $y>20$ for $N=40,80$ and 160
are shown in the right panel.}
\end{center}
\label{fig5}
\end{figure}

\section{Quasi-classical approximation. AA scattering}
Now we proceed to the general case of $AA$ scattering illustrated in Fig. 1$b$.
Unlike the fan case the action in the general can only be found by numerical methods.
One may consider two different techniques to this aim. The older method proposed in ~\cite{amati}
used evolution in rapidity of one of the fields (say $u$) down from $y$ where it is equal to $g_2$
according to the first Eq, (\ref{eqm}) in which $v$ is expressed via $u$ and the total energy $E$, which does not change
during evolution. Afterwards one has to determine $E$ from the condition $v(0)=g_1$. This method proved to be convenient for
analytic approach employed in ~\cite{amati} but turns out to be cumbersome and restricted to rather low values of the coupling constant.

A much more efficient method was used in our (also rather old) iterative method in which one solved the equations
of motion  (\ref{eqm}) using the mixed field from previous iterations.
In more detail,
one starts  evolving $u$ from its value $g_2$ at $y$ down to zero
taking $v=0$ in the mixed term
and $v$ from its value $g_1$ at rapidity zero  up to $y$ taking $u=0$ in the mixed term.
This gives our initial
solution $u_0$ and $v_0$. Then iterations are repeated taking the admixed  field
from the previous iteration, that is, according to
\[\dot{v}^{(n+1)}=\mu v^{(n+1)}-\lambda {v^{(n+1)}}^2-2v^{(n+1)}u^{(n)},\]
\[ \dot{u}^{(n+1)}=-\mu u^{(n+1)}+\lambda {u^{(n+1)}}^2+2\lambda u^{(n+1)}v^{(n)}.\]
Iterations are stopped when  the resulting $u$ and $v$ stop changing.

In fact in our method to find the entropy we need to know terms in the action
of a given power of the coupling constant $g=g_1=g_2$. To do this we develop $u$ and $v$
in powers of $g$, the dependence on them  following from the initial conditions:
\beq
u=\sum_{N=1}g^Nb_N,\ \ v=\sum_{N=1}g^N a_n\label{expuv}\eeq
(with $a_0=b_0=0$). The evolution equations transform into the system of evolution equations for
$a$ and $b$:
\beq
\dot{a}_N=\mu a_N-\lambda\sum_{M=0}^Na_Ma_{N-M}-2\lambda\sum_{M=0}^Na_Mb_{N-M},\label{eqa}\eeq
\beq
\dot{b}_N=-\mu b_N+\lambda\sum_{M=0}^Nb_Mb_{N-M}+2\lambda\sum_{M=0}^Nb_Ma_{N-M}.\label{eqb}\eeq
These equations are to be solved at each step of our iterational procedure.
At the first step one takes $b_n=0$ in (\ref{eqa}) and $a_n=0$ in (\ref{eqb}).
At the following steps one takes $b_n$ in (\ref{eqa}) and $a_n$ in (\ref{eqb}) from the previous iteration.

{\bf Numerical results}\\
We applied  the iterative procedure to find the action for $\mu=0.1$ and $\lambda=0.01$, which are
the popular values extracted from the comparison with the data. Taking the symmetric case with
$g_1=g_2=1$ we obtained the following values for action $\cal {A}$ at $y=10\div 50$
\beq
{\cal A}(y)=2.052,\,3.569,\,5.406,\,7.400,\,9.413,\ \ {\rm at}\ \ y=10,\,20,\,30,\,40,\,50.
\label{action}
\eeq
We do not present more detailed values for the action. since we are actually not interested in them
as a whole. To determine the probabilities $P_n$ to find $n$ pomerons we split the total action in parts
proportional to $g^N$ where $g=g_1=g_2$ is the common coupling constant to both nuclei to get
\beq
{\cal A}=\sum_{N=2}{\cal A}_Ng^N,\ \ {\cal A}_N=\lambda\int_0^Ydy \sum_{M=0}^Nb_M(y)\sum_{L=0}^{N-M}a_L(y)a_{N-M-L}(y) +b_{N-1}(0)
\label{expn}
\eeq
where $a_N$ and $b_N$ are the coefficients in (\ref{expuv})  obeying Eqs. (\ref{eqa})) and (\ref{eqb}).

The system of equations (\ref{eqa}) and (\ref{eqb}) was solved iteratively for $N=1,2,... N_{max}$ as described above. We chose $N_{max}=40$.
Recalling that the number of pomerons is $n=2N-3$ this allows to find the probabilities for the maximal
number $n_{max}=77$ pomerons.

As expected the solution of (\ref{eqa}) and (\ref{eqb}) gave ${\cal A}_N$ violently oscillating with $N$ with the amplitude
growing fast with rapidity and achieving values of the order $10^{70}$ for $N=40$ and $Y=40$. Summation
over $N\leq N_{max}$ gives absurd values for action, which is to be expected, since action is not at all
analytic in $g$. However the results allow to employ our procedure to extract the probabilities described in Section 3.
The values of the pomeron exchange found in this way grow with $y$ as $e^{\mu y}$ and the probabilities $P_n$
go down with $n$ very fast. In Fig. 6 we show these probabilities at three typical rapidities 7.5, 15 and 22.5.
The logarithmic scale illustrates the fast diminishing of the probabilities with $N$.
\begin{figure}
\begin{center}
\epsfig{file=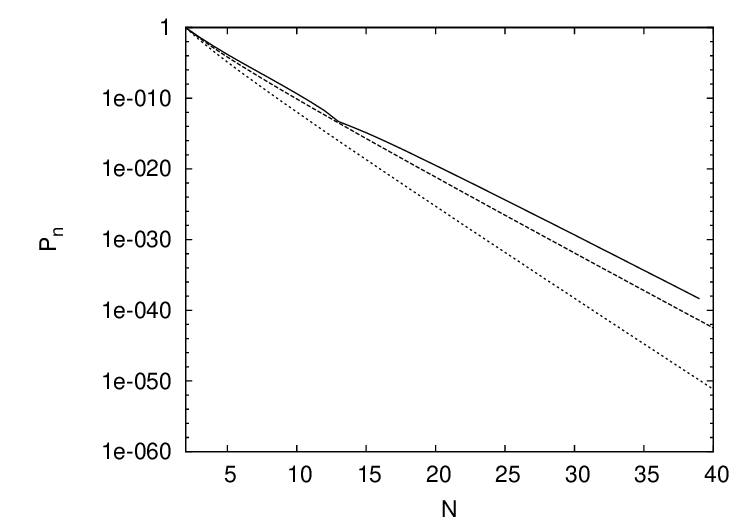, width=8 cm}
\caption{Probabilities $P_n$ to find $n=2N-3$ pomerons as a function of $N$ for the general quasi-classical case with $\mu=0.1$ and $\lambda =0.01$.
The curves from up to bottom correspond to $y=7.5,\ 15$ and 22.5}
\end{center}
\label{fig6}
\end{figure}

The resulting entropy is shown in Fig. 7 for $y=0\div 40$.  The AA entropy is found to be significantly higher than for fans, but the overall trend with rapidity is the same.
It rises achieving its maximal value at $Y\sim 7$ and then slowly falls at higher $Y$ going to zero at $Y>>1$ and so demonstrating vanishing
entanglement at high rapidities.

\begin{figure}
\begin{center}
\epsfig{file=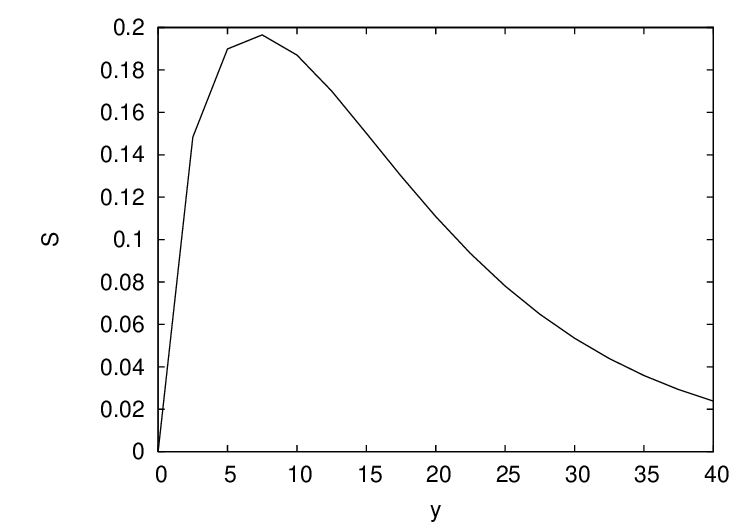, width=8 cm}
\caption{Entropy for the general quasi-classical  amplitude with $\mu=0.1$ and $\lambda =0.01$ as  a function
of rapidity}
\end{center}
\label{fig7}
\end{figure}


\section{Full quantum model}
\subsection{Reaction-diffusion approach}

In ~\cite{redif} an attempt was made to introduce the probability to create pomerons in the same way as for the fan diagram
approximation. Namely starting from the same expansion (\ref{powerexp})
they got a system of evolution equation for $c_n$ generalizing (\ref{cny})
\beq
\frac {dc_n(y)}{dy}=\mu n c_n(y)-\lambda (n-1)c_{n-1}(y)+\lambda n(n+1)c_{n+1}.
\label{cny1}
\eeq
Introducing then $\nu_n$  and $P_n$ by the same Eqs. (\ref{cn}) and (\ref{nun}) they obtained a system of equations for $P_n$,
which generalized Eq. (\ref{prob})
\beq
\dot{P}_n(\bar{y})=-nP_n(\bar{y})+(n-1)\rho_{n-1}(\bar{y})+\lambda (n+1)(n+2)\rho_{n+2}(\bar{y})-\lambda n(n+1)\rho_{n+1}(\bar{y}),
\ \ n=0,1,2,...
\label{prob1}
\eeq
They tried to interpret $P_n$ satisfying this equation as the probabilities to find $n$ partons in the quantum case.
However they  commented that this immediately met with the difficulty to guarantee  positiveness of $P_n$
necessary for such interpretation although they assured that this was irrelevant for their particular purpose in ~\cite{redif}.

To see  the problem and estimate its scope we evolved $P_n(y)$ according to Eq. (\ref{prob1}) for $\mu=0.1$ and $\lambda=0.01$
up to $y=30$ staring from $P_n(0)=\delta_{n1}$, which corresponds to pA scattering (now with loops).
 Our results depend on the cutoff $n<N$ in Eq. (\ref{prob1}).

Generally at all $y$ with the growth of $n$ $P_n$ first diminish smoothly until  a certain $n$ when they start
oscillating around zero. With the growth of $y$ these oscillations start earlier and their amplitude grows, so that
the interpretation of $P_n$ becomes impossible. This is illustrated in Fig. 8 with $n\leq N=80$.
Oscillations begin at $n=77$ when $y=9$, at $n=45$ when $y=18$ and already at $n=9$ when $y=27$ with the amplitude of subsequent oscillations
growing with $n$ and $y$. At $n=80$ this amplitude grows from $10^{-18}$ to 100 and $10^{30}$ at $y=9,18$ and 30 respectively.
If one takes greater cutoff $N=160$ then at $y=9$ the probabilities do not change until $n=80$. Afterwards they continue to
smoothly diminish  staying above unity until $n=160$ when they start oscillations with the amplitude of order $10^{-34}$.
At higher $y$ the situation worsens. At $y=18$ oscillations start already at $n=15$ with the amplitude around unity and at still higher $y$
the probabilities loose any sense with oscillations at all $n$ with  ever growing amplitudes reaching order $10^{130}$.
As a conclusion, one can interpret $P_n$ as a probability only at low rapidities $y<10\div 15$. At higher rapidities this interpretation is possible
up to certain $n=1,2,..n_{max}$ determined by  staying inside  interval $[0,1]$..

Accordingly we tentatively calculated the entropy  summing over only the states $n$ for which $0\leq P_n\leq 1$. The results are shown
in Fig. 9 for different cutoffs $N=40$, 80 and 160. As one observes  a reasonable entropy is found only at $y\leq 17$, when it turns out to be
 independent of the cutoff.
At higher $y$ the entropy  strongly depends on the cutoff and is not trustworthy.
\begin{figure}
\begin{center}
\epsfig{file=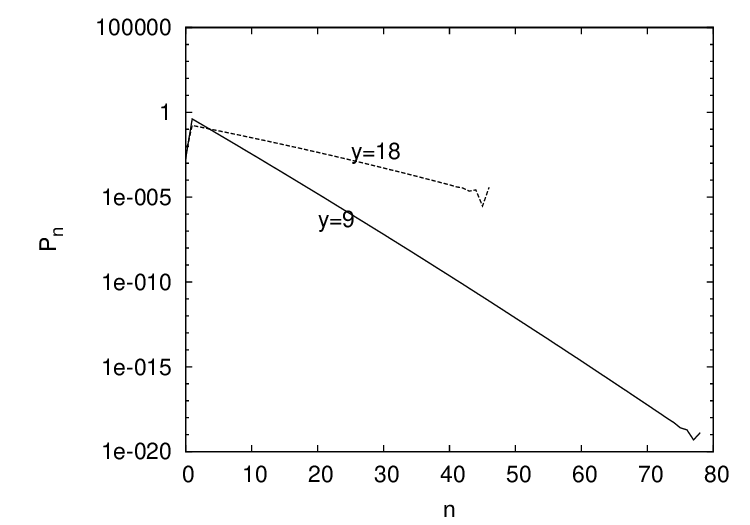, width=8 cm}
\epsfig{file=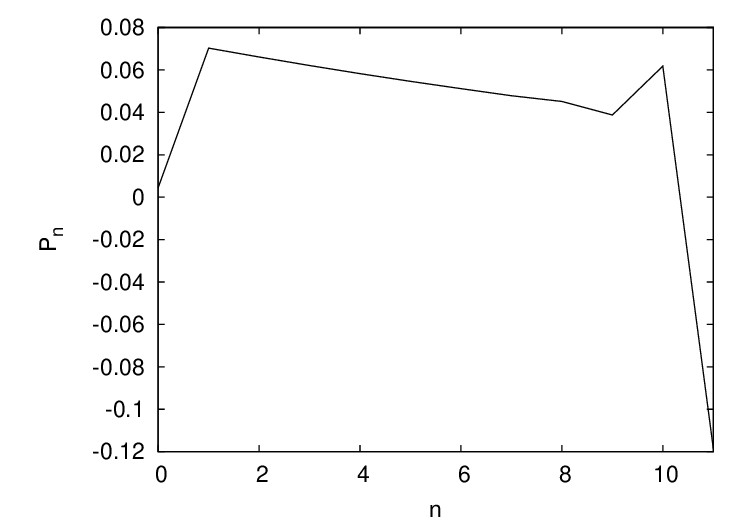, width=8 cm}
\caption{Probabilities $P_n$ from the reaction-diffusion approach with cut $n<80$ in the evolution equation for $P_n$.
at $y=9$ and 18 (left panel) and $y=27$ (right panel)}
\end{center}
\label{fig8}
\end{figure}

\begin{figure}
\begin{center}
\epsfig{file=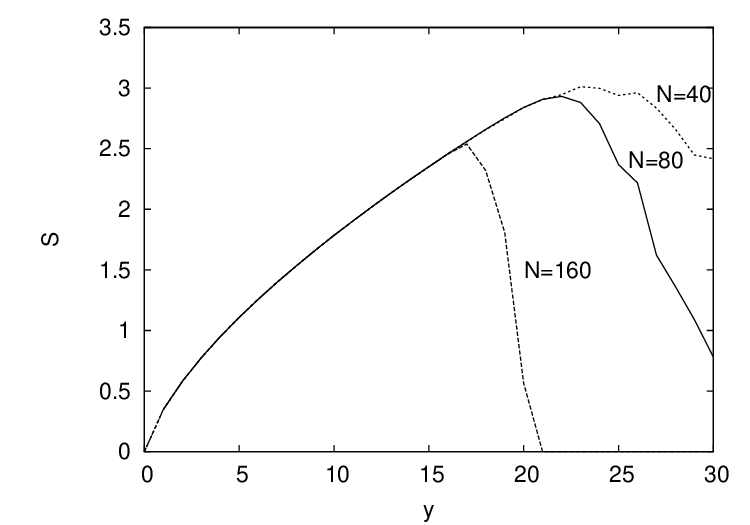, width=8 cm}
\caption{Entropy from the reaction-diffusion approach with cuts $n<N$ in the evolution equation for $P_n$.
Summation over $n$ restricted to its values  for which $0\leq P_n \leq 1$}
\end{center}
\label{fig9}
\end{figure}
\subsection{Rescaling method}
Unlike the reaction-diffusion approach our rescaling method introduced for the quasi-classical case in Section 4.2
is perfectly applicable also for the quantum case. One solves the system of equations (\ref{cny1}) for $c_n(y)$
and introduces the probabilities $P_n$ according to (\ref{rhoresc}) and (\ref{norm}):

\begin{center}
{\bf pA scattering}\\
\end{center}
For pA scattering one takes at $y=0$
$
\Psi(0,u)=g_1u$.
As in the previous section the amplitude then generalizes the fan diagram case to include loops.
We performed  numerical evolution  using   (\ref{cny1}) from $c_n(y=0)=\delta_{n1}$ with a cutoff $n\leq N$.
Again we took  $\mu=0.1$ and $\lambda =0.01$.
It turned out that a reliable solution could be found only at rapidities $y<20$
when it  practically does not depend on the cutoff.
At higher rapidities one observes
strong dependence of the cutoff, so that the solution could not be considered as reliable. This is illustrated in  Fig. 10
in which the probabilities found by our procedure are shown at $y\leq 20$ and $y\geq 20$ in the left and right panels respectively.
As a result the entropy corresponding to the calculated probabilities is uniquely found for $y<20$ but has widely different values
for different $N$ at
higher rapidities as shown in Fig. 11 together with the quasi-classical entropy (fans). As one observes the influence of
quantum corrections is quite small at $y<20$ but grows in an uncontrollable way at greater rapidities.

In conclusion, our recipe allows to construct probabilities $P_n(y)$ to find $n$ pomerons at a given $y$ as long as
$y$ is not so high, $y<20$. At higher $y$ coefficients $c_n(y)$ in the wave function cannot be found in a reliable manner. Then
the solution of their evolution system (\ref{cny}) heavily depends on the chosen cutoff $N$. Actually at large $n$ and $y$ the
coefficients $c_n(y)$ oscillate with  ever growing amplitudes reaching very high orders of magnitude and so going beyond any
reasonable precision in calculations.

\begin{figure}
\begin{center}
\epsfig{file=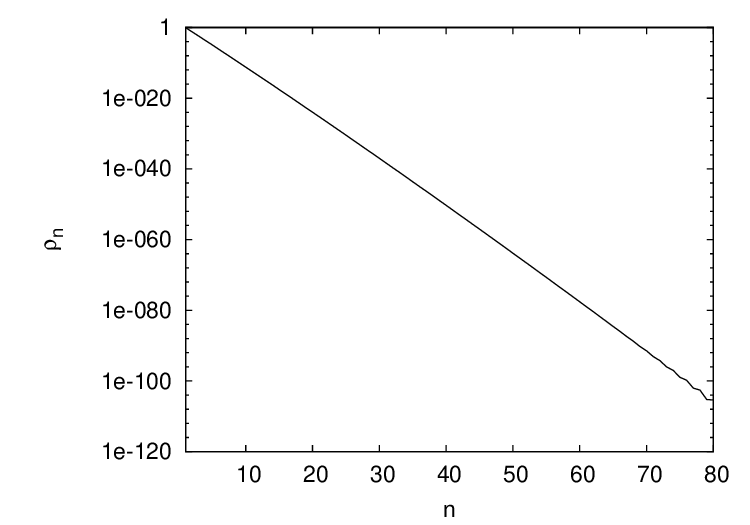, width=8 cm}
\epsfig{file=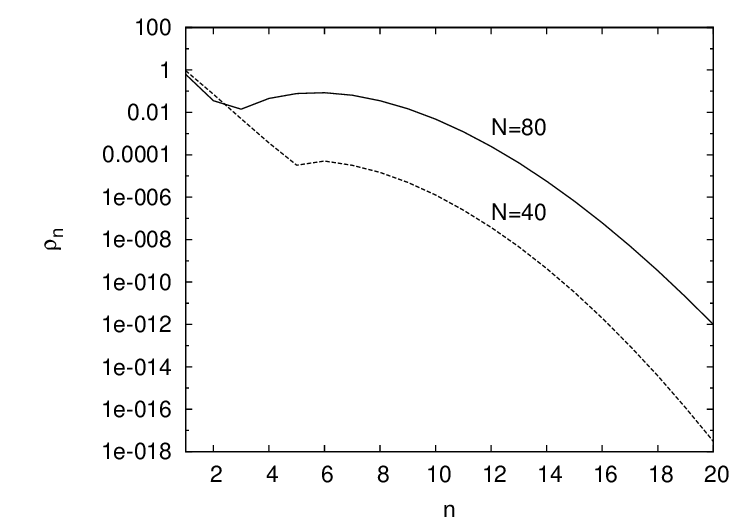, width=8 cm}
\caption{Probabilities $P_n$ for pA scattering from the rescaling method with  cutoff $n<N$ in the evolution equation for $P_n$.
at $y=10$  (left panel) and $y=27$ (right panel)}
\end{center}
\label{fig10}
\end{figure}

\begin{figure}
\begin{center}
\epsfig{file=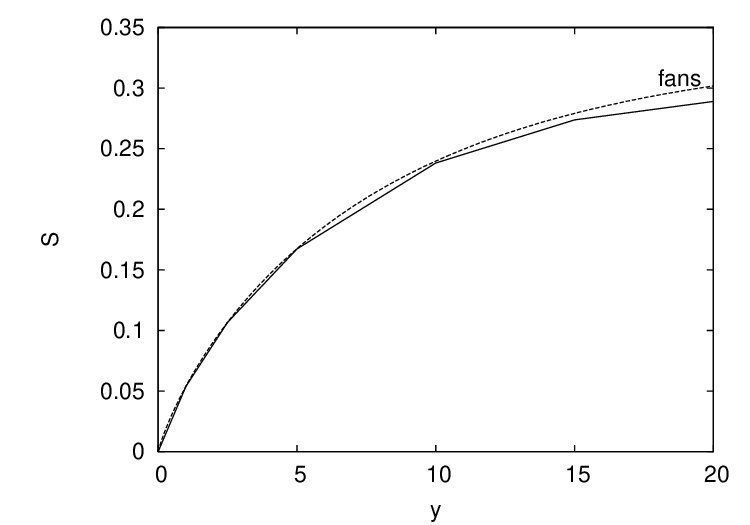, width=8 cm}
\epsfig{file=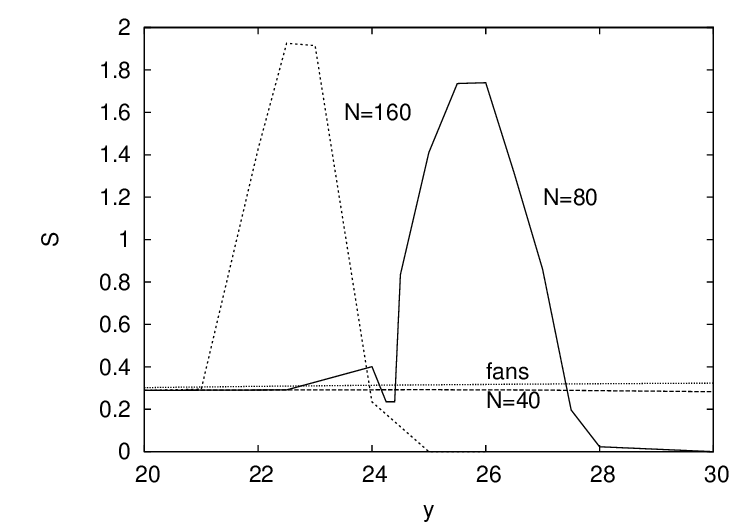, width=8 cm}
\caption{Entropy for pA scattering from the rescaling method with cutoff $n<N$ in the evolution equation for $P_n$
for $y=20$, independent of $N$, (left panel) and $y>20$ for $N=40,80$ and 160 (right panel)
Curves denoted as 'fans' show the quasi-classical entropy (\ref{sresc})}
\end{center}
\label{fig11}
\end{figure}

\begin{center}
{\bf AA scattering}\\
\end{center}
For AA scattering the initial state is given by (\ref{psi0}). So, in contrast to pA scattering,  already at $y=0$ the system  is a superposition of states
with different numbers of pomerons. In this case, putting $g_1=1$
$
c_n(0)=-(-1)^n/n!$.

Acting according to our rescaling procedure we take the probabilities at $y=0$
$
P_n(0)=R^{-n}(0)/n!$
and from the normalization condition find $R(0)=1/\ln 2$. So at $y=0$
the probabilities are
\beq
P_n(0)= \frac{(\ln 2)^n}{n!}.
\label{pa0}
\eeq
They strongly decrease as $n$ grows.
The corresponding entropy at $y=0$ is naturally greater than zero.
One finds
\beq
S(0)= 0.812166.
\label{sa0}
\eeq

With the growth of rapidity the probabilities and entropy are to be found from the system of evolution equations (\ref{cny1}) with the
initial conditions $c_n(0)$. As before they depend on the cutoff $n<N$ in this system.
Our results for the entropy are shown in Fig. 5 separately for $y<15$ and $y>15$.
Similarly to the hA case reliable results, not depending on $N$, can only be obtained at $y<18$, when the entropy grows monotonously from its value
0.812 at $y=0$ to  around unity at $y=18$. At higher rapidities our results strongly depend on the cutoff and actually reflect a poorly controllable
behavior of coefficients $c_n(y)$, which oscillate with ever greater amplitudes. To see it one can look upon the behavior of the scaling factor $R(y)$.
At $y<18$ it stays in the interval $1<R<8$ independent of $N$. But at greater $y$ it abruptly blows up reaching values $10^7$ at $y=26$ for
$N=40$ and $10^6$ already at $y=22$ for $N=80$. Such values lie outside the precision admitted in our evolution program.

\section{Conclusions}
We have studied the standard Regge-Gribov model with triple pomeron interactions from the point of view of the
probabilistic interpretation, which has long been the subject of discussion. We draw attention to the fact that
introduction of probabilities within this model is not unique
and depends on what is meant under the relevant substructures,
 We have considered three different choices of these
probabilities, namely the traditional partonic one, related to the reaction-diffusion approach
and based on the partonic substructure of the pomeron, a pomeronic one based on the representation
of the wave function in terms of multipomeron components and an alternative pomeronic one based on the number of pomeron propagators in the
relevant Feynman diagrams. The three sorts of probabilities are very different and, which is more important, behave very differently with
the growth of energy. Their entropy either grows roughly linearly with rapidity or saturates and tends to a
finite constant at large rapidities or, finally, first grows but then achieves a maximum and afterwards  diminishes to zero as rapidity grows.
Possible observable manifestation of these probabilities and entropy are also very different.
While partonic quantities are supposedly to be seen in the spectra of emitted hadrons, the pomeronic ones are rather to be seen in the
distributions of the cross-section in powers $n$ assuming that their dependence of the coupling constants $g$ to the participants
is presented as a series in $g^n$.
The partonic distributions in the probabilistic approach of ~\cite{musal,redif, bravac,kharzeev} seem to be external to
TRGM and are imported into it from the QCD. They actually refer to the zero order in perturbation series in TRGM, that is to the
single pomeron exchange.
As a result conclusions in ~\cite{kharzeev}  about  reaching an entangled state and scrambling at high energies
do not seem to be  directly related to our model but rather to the QCD interpretation of the pomeron in it.

\begin{center}
{\bf Acknowledgment}\\
\end{center}

\noindent The author is thankful for M.Lublinsky for numerous and helpful discussions.


\end{document}